# Delay Mitigation in Air Traffic Flow Management

Mehran Makhtoumi
Department of Physics
Universitat Politècnica de Catalunya, UPC
Barcelona, Spain
mehran.makhtoumi@upc.edu

***Abstract*— To mitigate ATFM delay, different approaches have been proposed so far which can be categorized into strategic and tactical domains. The strategical techniques mainly concern airport slot allocation and for the tactical domain, the ATFM function has several solutions available that range from the ground and air holding to rerouting actions, which have not gained significant efficiency in ATFM delay mitigation due to the fact that delays become apparent only on the tactical level when the strategic flight plan has been filled already. To tackle and address this problem there is a need for an algorithm that can synchronize strategical and tactical schedules. To fill this gap, in this paper the concept of fair buffer scheduling is proposed which can potentially contribute to strategical and tactical operations synchronization that would result in ATFM delay mitigation by increasing the system's robustness. The objective is to obtain an optimum fair and efficient buffer choice that mitigates ATFM delay and increases the stakeholders' welfare. Each appropriate and efficient approach requires a comprehensive understanding of the strategical buffer scheduling. This study presents a delay cost and flight buffer model that could be used for generating optimal buffer times to be considered as the initial population for the optimization problem to investigate the viability of employing fairness measures to obtain schedules with different trade-offs between cost, delay, and fairness.**

***Keywords—Delay, ATFM, Fairness, Buffer.***

***Nomenclature***

|  |  |
|---|---|
| $\emptyset$ | Binomial variable |
| $b_f$ | Flight buffer time |
| $\boldsymbol{f^t}$ | Scheduled flight time |
| $C_{f_{(1-P)}}$ | Flight earliness cost |
| $C_{f_{(P)}}$ | Flight lateness cost |
| $\alpha$ | Profit gained from a flight |
| N | Total income from a flight |
| $c_r$ | Gates rental costs |
| $T'_{S_G}$ | Scheduled ground times |
| $d_g$ | Ground delay |
| $G^t_{P\,(\emptyset)}$ | Ground duration probabilities |
| $b_g$ | Ground buffer times |
| $G^t$ | Ground duration |
| $C_{G_{(1-P)}}$ | Ground earliness cost |
| $C_{G_{(P)}}$ | Ground lateness cost |
| $C_{f(\emptyset)}$ | The expected value of a binomial variable for earliness and lateness |
| $c_{APU}$ | APU fuel consumption cost |
| $c_g$ | The ground or taxiing fuel consumption cost |
| $E'_{S_G}$ | Scheduled taxiing fuel consumption rate |
| $S_{cl}$ | Climb segments |
| $S_{cr}$ | Cruise segments |
| $S_{ds}$ | Descent segments |
| $T_{S_f}$ | Time spent on segments of a flight |
| D | Segment distance |
| V | Segment corresponding velocity |
| $T'_{S_f}$ | Scheduled flight time |
| $T''_{S_f}$ | Actual flight time |
| $F_S$ | Segment-associated fuel consumption |
| f | Fuel burn rate |
| $\tau_s$ | Engine thrust level |
| $c_\tau$ | Engine thrust level-specific fuel consumption |
| $E'_{S_f}$ | Scheduled engine flight fuel consumption |
| $E''_{S_f}$ | Actual flight fuel consumption |
| $c_f$ | Fuel cost per unit |
| $C_E$ | Flight delay fuel cost |
| $c_c$ | Crew cost of service |
| $c_p$ | Passenger cost |
| $c_m$ | Aircraft maintenance cost |
| $c_d$ | Aircraft depreciation cost |
| $d_f$ | Flight delay |
| $C_T$ | The total cost of flight delay |

## I. INTRODUCTION

Commercial aviation has evolved into a complex distributed transport system that interacts with precious resources, imbalanced demand, and a complicated origin-destination matrix that requires synchronization to function fluently. In this complex system, any disruption could affect subsequent flights which lead to the airport or airspace congestion, queues, and flights delay.

Delay is mentioned as the most significant performance indicator of transport systems, commercial aviation stakeholders consider delay as the period by which a flight is late. Delay can be defined by the difference between scheduled and real times of departure or arrival. Flight delays negatively affect passengers, airlines, crew, and airports and the main impact for the stakeholders is economic. The U.S. economy endures an estimated $33 billion annually as a result of the flight delays cost and in the EU this value is €11.5 billion.

An ideal delay cost model would require taking into account operational expenses such as crew, aircraft maintenance, aircraft depreciation, passenger, engine fuel consumption, and environmental expenses like noise, carbon dioxide, and nitrogen oxide emissions. Moreover, the marketing strategies of airlines are highly related to delays as customers' loyalty jeopardizes by delays since passengers prefer reliable operation. Thus delay levels, fares, flight frequency, and complaints about airline service are strongly interconnected, hence mitigation of ATFM delays is of great importance for all the stakeholders.

This paper includes six chapters in which it is explained how to fulfill the foregoing objectives. In the current chapter, a brief introduction is presented, and the previous investigations relevant to each of the objectives are surveyed in the second and third chapters. Afterward, the delay cost model is detailed in chapter four, which clarifies the cost variables and stakeholders. Later on, the flight buffer model is derived by defining and implementing the expected value of the binomial variable. Finally, the conclusions and a roadmap for further investigations are summed up in the sixth chapter.

## II. Delay modeling

ATFM delay incorporates both airports and en-route delays which is the consequence of new delays (en route), propagation delays, and cancellations. Therefore in the field of modeling delay, each research is based on the aforementioned domains upon their emphasis, hence to treat the problem of delay and delay cost modeling, various models have been proposed which can be divided into five categories: operational research, statistical analysis, machine learning, probabilistic models, and network representation.

### A. *Operational research*

Operations research is referred to the application of advanced analytical techniques in the pursuit of improved decision-making and efficiency which can range from simulation, mathematical optimization, and queueing theories to stochastic-process models, neural networks, and expert systems. In this regard for attaining a method that can be utilized in a dynamic setting, Soomer and Frnax [24] introduced a scaling approach for delay cost and provided a formulation for the flight landing time assignments implemented by a problem-specific local search heuristic which could generate robust schedules. To generate more robust aircraft and crew schedules, Duck et al. [25] presented a stochastic model for aircraft and crew scheduling to compute the occurrence of severe delays which leads to presenting an indicator for the stability of airline crew and aircraft schedules. Once more in a dynamic setting, Wieland [26] presented a delay model based on queuing theories by developing a tool to explore delay and the predictability of delay. The classical queuing concepts were also used to develop an intermediate queuing scenario by Kim and Hansen [27] to simulate counterfactual delay by applying an empirical and stochastic methodology. Furthermore, Pyrgiotis et al. [28] presented an analytical queuing and network decomposition method to model the propagation of delays within a large network of major airports and to compute the delays due to local congestion at individual airports.

### B. *Statistical analysis*

The statistical analysis for the delay and delay cost modeling is based on econometric, correlation, and regression models while multivariable analysis, parametric, and non-parametric tests are also utilized. The econometric specification of delays is presented by Morrison et al. [1] to assess the efficiency of air traffic control operations by developing an empirical delay model. Zou and Hansen [2] contributed to the investment benefit analysis methodology by proposing an assessment framework that incorporates a shift in the supply-demand system equilibrium to provide a delay model for excessive delay mitigation. In the case of applying binary choice models to derive utility functions and for flight cancellation motives, Xiong et al. [3] presented a model considering delay, size of aircraft, pre-existing delays, distance, frequency, and fare with airport and airline characteristics to model flight delay behavior and potential delay savings. The econometric analysis of A-check maintenance delay is presented by Mofokeng and Marnewick [4] to find the factors that contribute to delays during A-check maintenance, whereas Qin and Yu [5] used econometric modeling to analyze the periodicity of flight delay rate to figure out the influence of time factor on flight delays. Econometric modeling is also utilized to understand secondary delay generation by Hao et al. [6] who proposed a model that explains delay propagation behavior to other airports caused by an original airport delay. The non-parametric directional output distance function is utilized by Pathomsiri et al. [7] to present a statistical inference approach for the evaluation of airports' efficiency concerning delayed flights. Besides these, correlation analysis was derived by Abdel-Aty et al. [8] to figure out the major sources of delay for identifying the periodic patterns of arrival delay for non-stop domestic flights while Reynolds et al. [9] detected a correlation between delays and airport capacities to assess the capacity of the EU's airport infrastructure and the factors determining capacity.

### C. *Machine learning*

Estimation of system behaviors comprehended from provided data and derived by an algorithm is the aim of machine learning. In this regard, the research addressing the issue of delay is extending to machine learning which encompasses fuzzy logic, neural networks, K-nearest neighbor, and random forest methods. The random forest method has been used by Rebollo and Balakrishnan [15] to model root delay which considers both temporal and spatial delay states as explanatory variables that represent the system as a whole. A reinforcement learning algorithm has been utilized by Balakrishna et al. [16] to model taxi-out delays for effectively managing fuel consumption, emissions, and cost. Zonglei et al. [17] presented a method to model delay grades from the unsupervised flight data utilizing a k-nearest neighbor algorithm. Furthermore, an adaptive network-based approach was applied by Khanmohammadi et al. [18] to model flight delay for appropriately capturing the dynamic behavior of the aircraft landing problem.

### D. *Probabilistic modelling*

The probability estimation of an incident derived from previous data is referred to as probabilistic modeling, numerous probabilistic techniques have found their application in delay and delay cost modeling. In this sense, Mueller and Chatterji [10] by using density functions could model en route, arrival, and departure delays which recommend that Poisson distribution is the best candidate to estimate en route and arrival delays. Worth noting that a combination of genetic algorithm with the expectation-maximization approach is used by Tu et al. [11] to present a departure delay model of an airport. Also, survival modeling is applied by Wong and Tsai [12] to model the departure and arrival delay by utilizing the Cox proportional hazards model. Boswell and Evans [13] could utilize the probabilistic mass function to model delay and its propagation via a transition matrix and analyze the probability to cancel a flight when its previous flight is delayed. Furthermore, Monte Carlo simulations have attained attention in this field where Zhong et al. [14] used this method to model airports' runway capacity concerning flight delays.

### *E. Network representation*

The analysis of the delay in a graph theory is referred to as network representation. As pioneer research, AhmadBeyghi et al. [19] presented distinct flight schedules delay propagation trees for a range of initial lengths of delay to analyze the potential for delays and adequate slack possibilities to absorb flight delays in passenger airlines networks. Also, a directed acyclic graph with nodes and arcs is implemented by Abdelghany et al. [20] to present a flight delay projection model that projects flight delays, whereas the Bayesian network has been applied by Xu et al. [21] to investigate and visualize the delays among airports. In this regard, Wu et al. [22] developed a flight string-based Bayesian network model for delay modeling to indicate the impact of abnormal operating conditions. Further, Baspinar and Koyuncu [23] introduced a novel delay model by utilizing the epidemic spreading process to understand delay behavior over the network.

## III. Delay mitigation methods

ATFM delay not only affects negatively the stakeholders (e.g. airline, crew, and passengers) but also could cause environmental damage. Several methods exist to mitigate delay which are taxonomized as strategic and tactical approaches. In the following paragraphs, these methods are reviewed.

### *A. Tactical approaches*

Tactical approaches according to the US research literature refer to the planning after the flight take-off while in the EU this interval begins on the day of departure until take-off, these approaches imply techniques that impose flight delays to gain higher efficiency by utilizing ground delay procedures. In this regard, the pioneering work presented by Odoni [29] initiated the tactical flight planning framework, and several studies were established in this direction, the classification of these researches' methodologies range from single and multiple airport ground holding problems to ATFM problems and rerouting techniques. The objective of the ground-holding problem (GHP) is to implement ground-holding techniques to minimize the total cost of delay.

- **Single airport GHP:** The single airport GHP is considered a convenient planning technique that is capable of including several take-off and landing constraints to achieve the optimal plan of an airport. The fact that the single airport GHP assumes arrivals and departures as independent variables leads to a simple model since at congested airports those elements are highly dependent.

  The early efforts in this direction are detected in the studies by Andreatta et al. [30] where an extensive set of computational experiments are applied to the different single airport GHP models, further, the static and dynamic optimal solutions were tested by Richetta [31] using stochastic linear programming to assign the ground holds. Bianco et al. [32] proposed a combinatorial optimization approach obtained by a pseudo-enumerative procedure for aircraft sequencing problems. Hoffmann and Ball [33] could develop different models for the single-airport GHP considering banking constraints within fixed time windows to turn aside the delay propagation throughout the system. The problem of finding the optimal trade-off between the number of arrivals and departures was addressed by Dell'Olmo and Lulli [34] to minimize a delay function of total flights by implementing a more realistic representation of the airport capacity. Ma et al. [35] presented a mathematical model for GHP based on multi-commodity dynamic network flow for short-term ATFM to tackle capacity decrease due to sudden bad weather circumstances.

- **Multi airport GHP:** Multi airport GHP deals with multiple airports interaction aiming to present an optimal plan for satisfying the different airports' constraints and the fundamental assumption is that the departure and arrival capacity of the airports is deterministic functions of the time while considering unlimited capacity in the air sectors and avoiding alternative routes or flight speed from the model. The first attempt to address a multi-airport GHP in a dynamic environment where presented by Vranas et al. [38] in which several algorithms were proposed to update ground-holding decisions using pure IP formulations to gain the flexibility of extending to various degrees of modeling details, later, Andreatta and Tidona [39] could extend the previous formulations presented by Vranas et al. [38] to increase the efficiency of experimental implementation in terms for reducing computation time and optimality of solutions. In this regard, Brunetta et al. [36] solved GHP for a multi-airport problem by proposing a new algorithm based on priority rules where the flight priority is computed as a cost function, likewise, priority rules also utilized by Jiang et al. [37] to build a multi-airport GHP model for minimizing the total cost. Also, a multi-airport GHP model is developed by Zhang et al. [40] implementing a co-evolutionary genetic algorithm to minimize the cost of both airborne and ground holding.

- **Air Traffic Flow Management Problem (ATFMP):** This method aims to take into account a wide range of techniques and constraints on traffic flow management (TFM) problems for adapting dynamic air traffic demand with the accessible airport and airspace sectors capacities tactically to minimize delay and consequently delay costs, in this regard Odoni [41] spelled out key conceptual framework in ATFMP where emphasized on the configuration of participatory decision-making system for airspace operators along with stakeholders and further disclosed concerns for distributive effects of resource allocation beneath the issue of fairness. The pioneer base model in addressing capacity restrictions on the en route airspace is provided by Bertsimas and Patterson [42] as linear programming with relaxation technique which supports different modeling variations for dependencies between arrival and departure capacities, hub connectivity with multiple connections, flights banks, and rerouting aircraft, this base model was later optimized by Rios and Ross [43] by applying Dantzig–Wolfe decomposition method to decrease computational time. The base model was also extended as integer programming in Koepke et al. [44] to solve the issue of maximum on ground in military air mobility network for delay allocation to aircraft on the ground for the sake of minimizing the effects of disruptions, moreover, the temporal availability of routes was investigated by Geng and Cheng [45] by developing an integer programming model.

  Respecting the ATFM slot allocation, Plusquellec and Manchon [46] proposed constraint programming to overcome the drawbacks and lack of efficiency in the Computer-Assisted Slot Allocation (CASA) system which uses a greedy algorithm to dynamically allocate regulated departure slots on a first come first served policy where the presented constraint programming in this research suffered unevenly distributed workload and capacity violation and to address this problem, Barnier et al. [47] provided constraint programming with sliding discrete windows and sorting model. Later, Flener et al. [48] used constraint programming to model and efficiently solve the problem of minimizing and balancing traffic complexities of the airspace of adjacent sectors.

*B. Strategical approaches*

Strategical approaches according to the US research literature refer to the planning before the flight take-off while in the EU this interval covers six months to some days before departure, these approaches mainly imply airport slot allocation procedure to achieve demand management strategically.

- **Strategic slot scheduling:** A slot is defined as a specific time interval in which an airline is allowed to utilize airport facilities. Strategic slot allocation is considered as the major solution to address the reduced capacity problem at the airports, which refers to the allocation of slots for departure and arrivals in a constrained environment. The allocation procedures are formed by administrative regulations and priority-based rules as in the cases of grandfather rights and use it or lose it which have been established gradually over time. The research for slot scheduling approaches is categorized as market-based or scheduling methods.
  Market-based approaches assign a value to slots taking into account the real market cost of access to congested airports. In this regard, an important guide for the formulation of congestion pricing rules has been provided by Brueckner [49] in which it is proposed that congestion tolls for different airlines be different from the airlines' share tolls. Mott MacDonald [50] introduced secondary trading mechanisms for runway slots at congested community airports to achieve economic impact assessments based on slot usage forecasts. Madas and Zografos [51] developed a methodological framework for a multi-criteria evaluation slot allocation strategy concerning policy criteria and priorities in various airport settings to examine potential acceptability from stakeholders' points of view.
  Scheduling methods optimize the strategic slot assignment for arrivals and departures at an airport in a given scheduling period. In this regard, Corolli et al. [52] presented two stochastic programming models for the allocation of time slots over a network of airports to gain a tradeoff between the schedule and request discrepancies which refers to the time difference between allocated time slots and airline requests, and operational delays. Castelli et al. [53] investigated the possibility of fair redistribution of costs due to the imbalance between demand and capacity among airlines through monetary compensations to remove grandfather rights without significantly penalizing airlines.

- **Buffer scheduling:** Buffer is defined as the extra or additional time added to the flight or ground time, denoted as flight buffer (i.e. block time buffer) and ground buffer, which aims to mitigate delays and costs. Schedule padding is essential for air carriers in order to find schedules that work with the typical patterns of delay so that they can deliver passengers on time, and get maximum use out of their aircraft. In other words, a buffer can be comprehended as the forecast amount of time inserted into a schedule which is equal to the time required for the activity plus an amount to sustain the service on time. In this regard, AhmadBeygi et al. [54] suggested ground buffers as the solution for the delay mitigation problem by constructing a delay tree structure to investigate how delay propagates to the downstream flights, moreover, this study indicates that morning delayed flights lead to more delay propagation therefore early flights of the day need to receive more buffer, further Churchill et al. [55] presented two models for their delay problem and accounted ground buffering as a measure for delay mitigation problem. Wong and Tsai [56] proposed models for departure and arrival delay with a sight to ground and flight buffers aiming to provide flexible and reliable schedules against disruptions to mitigate flight delays. The issue of reallocating ground buffer with rescheduling departures was presented by AhmadBeygi et al. [57] to mitigate delay and prevent no extra cost whereas the issue of robustness of schedules was investigated by Arikan et al. [58] with several metrics by applying different strategies in scheduling ground and flight buffer for delay mitigation purpose. The issue of setting additional flight buffer and the related cost of earliness has gained the attention of the researchers where Deshpande and Arikan [59] by analyzing the cost ratio of leverage to underage for lateness and earliness concluded that earliness costs are more than that of lateness. Worth mentioning that the econometric cost functions forecasting different delay-buffer models and scheduled block time impact cost are presented by Zou and Hansen [60], furthermore Cook and Tanner [61] demonstrated the fact that inserting flight and ground buffers enhance delay mitigation that comes at the increment of the strategic cost which leads aircraft operators to pick a balanced choice from the buffer and delay trade-offs. Ivanov et al. [62] have proposed an approach for assigning ATFM delay over the buffers, and Villemeur [63] has studied the buffer issue on two occasions, first when the airlines are the deciding for setting buffers and the second case where a social planner does the same job.

### *C. ATFM fairness metric*

Fairness or equity is a state in which a stakeholder's welfare is increased to the extent possible and the fairness metric is a criterion to judge scheduling and a process to do so. Spelling out a criterion has been highly controversial hence different approaches are presented to measure fairness in ATFM.

For instance in GDP, traditionally fairness is perceived as prioritizing flights on a first-come-first-served (FCFS) perspective which could lead to a large mean response time as in the case of a double penalty, further on Collaborative Decision Making (CDM) presented a way out from the previous perspective by presenting first-scheduled first-served (FSFS) policy comprising ration by schedule (RBS) and compression algorithms where priority is determined by a combination of the Official Airline Guide (OAG) and daily operational data. In this regard considering a GDP, Vossen et al. [64] have developed an approach that can be considered as a combination between RBS and a pure proportional approach where ideal allocation overtime is defined as RBS to provide an optimization model that produces an allocation with minimum deviation from ideal one, therefore the proposed model minimizes the squared deviations between slots allocated and their ideal one that is subjected to constraints. Bernhart et. al. [65] developed fairness metrics to measure deviation from FSFS policy which incorporates time order deviation approximation and ration by schedule exponential penalty models extended to a multi-resource setting, the developed models are consistent with FSFS and can be directly applied to minizine the metric. The benefits of an optimal scheduler incorporating fairness were investigated by Rios and Ross [66] in which the fairness model was implemented to the previously mentioned ATFM model by Bertsimas and Patterson [42] to investigate design choices of equity for stakeholders. In terms of increasing the efficiency of the GDP, Glover and Ball [67] have proposed a two-stage multiobjective approach incorporating a ration by distance algorithm which allows to balance equity and efficiency by considering a maximum and total deviation from RBS. Manley and Sherry [68] have taken into account passenger flow efficiency and developed a GDP rule simulator to calculate efficiency and equity metrics by implementing two new algorithms of ration by aircraft size and ration by the passenger. In a network setting, Bertsimas and Gupta [69] have generalized the RBS to propose a discrete optimization model that incorporates an equitable distribution of delays among airlines by introducing a notion of fairness in network ATFM models which minimizes the total amount of overtaking and reversals.

Jaquillat and Vaze [70] have addressed scheduling interventions and proposed new metrics for inter-airline equity to balance scheduling adjustments fairly among airlines and maximize jointly efficiency and equity which is attained by minimizing the first largest stakeholder disutility and then the second-largest, and so on, which refers to the MAX-MIN fairness approach.

## IV. DELAY COST MODELING

Any credible model needs to clarify all stakeholders to attain an efficient and accurate model, the delay cost for the three main stakeholders of the crew, aircraft, and passengers required to be taken into account. The variables considered are crew cost, aircraft maintenance cost, aircraft depreciation cost, passenger cost, and engine fuel consumption cost. The flight delay and engine fuel consumption models are derived to achieve an accurate model that comprises the aforementioned variables.

The flight delay and engine fuel consumption models are obtained when the climbing, cruising, and descending segments of flight are differentiated. Let climb, cruise, and descent segments be denoted as $S_{cl}, S_{cr}$ and $S_{ds}$ and can be divided into $a, b,$ and $c$ sections to preserve accuracy (i.e. for fuel consumption calculations), therefore segments of a flight $S_f$ is defined

$$S_f = S_{cl} + S_{cr} + S_{ds} = \sum_{i=1}^{a} s_{cl}^{i} + \sum_{j=1}^{b} s_{cr}^{i} + \sum_{k=1}^{c} s_{ds}^{i}$$

deriving the delay model is feasible by expressing the above variables as the function of time spent in each segment, the time spent on segments of a flight $T_{S_f}$ is

$$T_{S_f} = T_{S_{cl}} + T_{S_{cr}} + T_{S_{ds}} = \frac{D_{cl}}{V_{cl}} + \frac{D_{cr}}{V_{cr}} + \frac{D_{ds}}{V_{ds}}$$

where $D$ is the segment distance and $V$ is the corresponding velocity. Therefore the scheduled flight time $T'_{S_f}$ is

$$T_{S_f}^{1} = \sum_{i=1}^{a} \frac{D_{cl}^{i}}{V_{cl}^{i}} + \sum_{j=1}^{b} \frac{D_{cr}^{i}}{V_{cr}^{i}} + \sum_{k=1}^{c} \frac{D_{ds}^{i}}{V_{ds}^{i}}$$

and the actual flight time is defined as $T''_{S_f}$ where climb, cruise, and descent segments are divided into $d, e,$ and $f$ sections

$$T_{S_f}^{2} = \sum_{l=1}^{d} \frac{D_{cl}^{l}}{V_{cl}^{l}} + \sum_{m=1}^{e} \frac{D_{cr}^{m}}{V_{cr}^{m}} + \sum_{n=1}^{f} \frac{D_{ds}^{n}}{V_{ds}^{n}}$$

The fuel cost is extracted when the notion of Segment Specific Range (SSR) is considered which corresponds to the aircraft's Kilometer per Kilogram (km/kg) fuel consumption and is defined as the ratio between the segment and the associated fuel consumed

$$SSR = \frac{D_S}{F_S}$$

where $F_S$ also can be represented as

$$F_S = \frac{V_s}{f}$$

here $V_s$ denotes the velocity and $f$ is the fuel burn rate which is related to the engine thrust level $\tau_s$ and engine thrust level-specific fuel consumption $c_\tau$

$$f = \tau_s . c_\tau$$

substituting the equations to obtain the scheduled engine fuel consumption gives

$$E'_{S_f} = \sum_{i=1}^{a} \frac{D_{cl}^{i}}{V_{cl}^{i}} . (\tau_{cl} . c_\tau) + \sum_{j=1}^{b} \frac{D_{cr}^{i}}{V_{cr}^{i}} . (\tau_{cr} . c_{cr}) + \sum_{k=1}^{c} \frac{D_{ds}^{i}}{V_{ds}^{i}} . (\tau_{ds} . c_{ds})$$

likewise, the actual flight fuel consumption is

$$E''_{S_f} = \sum_{l=1}^{d} \frac{D_{cl}^{l}}{V_{cl}^{l}} . (\tau_{cl} . c_\tau) + \sum_{m=1}^{e} \frac{D_{cr}^{m}}{V_{cr}^{m}} . (\tau_{cr} . c_{cr}) + \sum_{n=1}^{f} \frac{D_{ds}^{n}}{V_{ds}^{n}} . (\tau_{ds} . c_{ds})$$

Once the engine fuel consumptions are available, assuming fuel cost per unit as $c_f$ the flight delay fuel cost $C_E$ is then

$$C_E = c_f (E_{S_f}^{2} - E_{S_f}^{1})$$

Alongside the delay fuel cost of the engine, the total delay cost requires that the other stakeholders be taken into account accordingly, elements like crew cost of service $c_c$, passenger cost $c_p$, aircraft maintenance cost $c_m$ and aircraft depreciation cost $c_d$ are assumed, if the flight delay $d_f$ is

$$d_f = T_{S_f}^{2} - T_{S_f}^{1}$$

then the total cost of the flight delay $C_T$ is denoted as

$$C_T = \left( (c_c + c_p + c_m + c_d)\, d_f \right) + C_E$$

## V. FLIGHT AND GROUND BUFFER MODEL

The flight buffer model is calculated by defining and implementing the expected value of the binomial variable, to achieve that, the binomial variable $\emptyset$ is defined for flight $f_i$ such that

$$\emptyset : \begin{cases} P(\emptyset = 1) = P \\ P(\emptyset = 0) = 1 - P \end{cases}$$

the probability of undisrupted flight ($\emptyset = 0$) is $1 - P$, and the probability of disrupted flight ($\emptyset = 1$) is $P$, therefore the flight duration probabilities $f_{P\,(\emptyset)}^{t}$ are

$$f_P^t = T_{S_f} + d_f$$
$$f_{(1-P)}^{t} = T_{S_f}$$

To tackle the possible harm caused by delays, flight buffer times $b_f$ are added to the scheduled flight time $f^t$ leading to $f^t = T_{S_f} + b_f$ then flight earliness time $FET$ for undisrupted flight $d_f = 0$ is

$$FET_{f_{(1-P)}} = f^t - f_{(1-P)}^{t} = T_{S_f} + b_f - T_{S_f} = b_f$$

for a disrupted flight when $d_f > b_f$ the flight lateness time $FLT$ can be written as

$$FLT_{f_{(P)}} = f_P^t - f^t = T_{S_f} + d_f - T_{S_f} - b_f = d_f - b_f$$

and for the case where $d_f < b_f$ the flight lateness time $FLT$ is denoted as

$$FLT_{f_{(P)}} = f^t - f_P^t = T_{S_f} + b_f - T_{S_f} - d_f = b_f - d_f$$

for the sake of the profit maximization process, the $b_f$ will be starting to choose increasingly from $b_f < d_f$, so the above $FLT$ equations can be rewritten as

$$FLT_{f_{(P)}} = f_P^t - f^t = d_f - b_f$$

To define the earliness cost $C_{f_{(1-P)}}$ which mainly impacts crew and passengers, the following expression is considered

$$C_{f_{(1-P)}} = (1-P)\,(c_c + c_p)\,\left(FET_{f_{(1-P)}}\right)^2$$

and the lateness cost $C_{f_{(P)}}$ is

$$C_{f_{(P)}} = (P)\left(c_c + c_p + c_m + c_d\right)\left(FLT_{f_{(P)}}\right)^2$$

Therefore the expected value of a binomial variable for earliness and lateness cost $C_{f(\emptyset)}$ is

$$C_{f(\emptyset)} = C_{f_{(P)}} + C_{f_{(1-P)}}$$

The aim is to maximize the profit of the airline concerning costs bounded by stakeholders, to gain further perception assume that profit gained $\alpha$ of an airline from a flight is stated as

$$\alpha = N - C_{f(\emptyset)} - \Big(c_f(E'_{S_f} + b_f) + (c_c + c_p + c_m + c_d)(T'_{S_f} + b_f) + (c_c + c_p + c_r)\,T'_{S_G}\Big)$$

where $N$ is the total income from a flight, $c_r$ is the gates rental costs and $T'_{S_G}$ is scheduled ground times, hence the profit maximization is the buffer choice that minimizes $C_{f(\emptyset)}$, to do so the first-order condition for flight buffer is obtained

$$2(1-P)\,(c_c + c_p)\,b_f - 2\,(P)\left(c_c + c_p + c_m + c_d\right)(d_f - b_f) - (c_f + c_c + c_p + c_m + c_d) = 0$$

which leads to the following expression for the flight buffer $B_f$

$$B_f = \frac{P\left(c_c + c_p + c_m + c_d\right)}{(1-P)(c_c + c_p) + P\left(c_c + c_p + c_m + c_d\right)}\,d_f + \frac{c_f + c_c + c_p + c_m + c_d}{2\left((1-P)(c_c + c_p) + P\left(c_c + c_p + c_m + c_d\right)\right)}$$

The ground buffer is the extra time added to scheduled ground times where $T'_{S_G}$ is equal to the minimum amount needed to cover the turnaround and taxiing times. Considering the aforementioned binominal variables for the flight $f_i$ such that

$$\emptyset: \begin{cases} P(\emptyset = 1) = P \\ P(\emptyset = 0) = 1 - P \end{cases}$$

the probability of undisrupted ground operation $(\emptyset = 0)$ is $1 - P$, the probability of disrupted ground operation $(\emptyset = 1)$ is $P$ and ground delay denoted as $d_g$ then the ground duration probabilities $G^t_{P\,(\emptyset)}$ are

$$G^t_P = T'_{S_G} + d_g$$
$$G^t_{(1-P)} = T'_{G_S}$$

To tackle the possible harm caused by ground delays, ground buffer times $b_g$ are added to the ground duration $G^t$ leading to $G^t = T_{S_G} + b_g$ then ground earliness time $GET$ for undisrupted ground operation $d_g = 0$ is

$$GET_{f_{(1-P)}} = G^t - G^t_{(1-P)} = T_{S_G} + b_g - T_{S_G} = b_g$$

for a disrupted ground operation when $d_g > b_g$ the ground lateness time $GLT$ can be written as

$$GLT_{f_{(P)}} = G^t_P - G^t = T_{S_G} + d_g - T_{S_G} - b_g = d_g - b_g$$

and for the case where $d_g < b_g$ the flight lateness time $GLT$ is denoted as

$$GLT_{f_{(P)}} = G^t - G^t_P = T_{S_G} + b_g - T_{S_G} - d_g = b_g - d_g$$

for the sake of the profit maximization process, the $b_g$ will be starting to choose increasingly from $b_g < d_g$, so the above $GLT$ equations can be rewritten as

$$GLT_{f_{(P)}} = d_g - b_g$$

To define the ground earliness cost $C_{G_{(1-P)}}$ which mainly impacts crew and passengers, the following expression is considered

$$C_{G_{(1-P)}} = (1-P)\,(c_c + c_p)\,\left(GET_{G_{(1-P)}}\right)^2$$

in continuous form as

$$C_{G_{(1-P)}} = (c_c + c_p)\int_{d_g}^{b_g} GET_{G_{(1-P)}}\,dd_g$$

and the lateness cost $C_{G_{(P)}}$ is

$$C_{G_{(P)}} = (P)\left(c_c + c_p\right)\left(GLT_{G_{(P)}}\right)^2$$

in continuous form as

$$C_{G_{(P)}} = (P)\left(c_c + c_p\right)\int_{b_g}^{d_g} GLT_{G_{(P)}}\,dd_g$$

Therefore the expected value of a binomial variable for earliness and lateness cost $C_{f(\emptyset)}$ is

$$C_{G(\emptyset)} = C_{G_{(P)}} + C_{G_{(1-P)}}$$

Recalling profit gained $\alpha$ of an airline from a flight is stated as

$$\alpha = N - C_{G(\emptyset)} - \Big(c_g\big(E'_{S_G}\big) + c_{APU}(b_g) + (c_c + c_p + c_m + c_d)(T'_{S_f}) + (c_r)(T'_{S_G})\Big)$$

where is $c_{APU}$ refers to Auxiliary Power Unit (APU) fuel consumption cost, $c_g$ is the ground or taxiing fuel consumption cost, and $E'_{S_G}$ is the scheduled taxiing fuel consumption rate. Hence the profit maximization is the buffer choice that minimizes $C_{G(\emptyset)}$, to do so the first-order condition for the ground buffer is obtained

$$2(1-P)\,(c_c+c_p)\,b_f-2\,(P)\big(c_c+c_p\big)\big(d_f-b_f\big)-(c_{APU}\,)=0$$

which leads to the following expression for the ground buffer

$$B_g=\frac{P\big((c_c+c_p\,\big)}{(1-P)(c_c+c_p)+P\big((c_c+c_p\,\big)}d_g+\frac{c_{APU}}{2\big((1-P)(c_c+c_p)+P(c_c+c_p)\big)}$$

## VI. CONCLUSION

In this paper, a comprehensive model for the delay cost is presented for three main stakeholders of the crew, aircraft, and passengers considering variables like crew cost, aircraft maintenance cost, aircraft depreciation cost, passenger cost, and engine fuel consumption cost. Using the delay cost model, the flight buffer model is derived by defining and implementing the expected value of the binomial variable, and profit maximization is used to determine the buffer choice that minimizes delay cost. This study proposes the concept of fair buffer scheduling which can potentially contribute to strategical and tactical operations synchronization that would result in ATFM delay mitigation by increasing the system robustness since previous solutions that range from the ground and air holding to rerouting actions, could not gain a significant efficiency in ATFM delay mitigation due to the fact that delays become apparent only on the tactical level when the strategic flight plan has been filled already. Using the presented models would enable the generation of optimal buffer times that would be considered as the initial population for the optimization problem. In addition, the ATFM delay model could be presented which requires considering the Network Manager (NM) allocation of ATFM slots to flights that produce ATFM slot allocation delay based on historical data. Once the ATFM delay model is attained, the ATFM delay mitigation model could be developed where the buffer time is incorporated to investigate the viability of employing fairness measures to obtain schedules with different trade-offs between cost (operational expenses), delay, and fairness.